\documentclass[a4paper,10pt,twocolumn,twoside]{cpc-hepnp}
\usepackage{multicol}
\usepackage{multirow}
\usepackage{graphicx}
\usepackage{booktabs}
\usepackage{amssymb,bm,mathrsfs,bbm,amscd}
\usepackage[tbtags]{amsmath}
\usepackage{lastpage}
\usepackage[english]{babel}
\usepackage{epsfig}
\usepackage{graphicx}
\usepackage{epstopdf}
\usepackage{graphbox}
\usepackage{xcolor}
\usepackage{hyperref}

\begin{document}

\fancyhead[c]{\small Chinese Physics C~~~Vol. xx, No. x (2024) xxxxxx}
\fancyfoot[C]{\small 010201-\thepage}
\footnotetext[0]{Received \today}

\title{Quantifying the memory and dynamical stability of magnetar bursts\thanks{Supported by the National Natural Science Fund of China (Grant Nos. 12005184, 12175192, 12275034 and 12347101), and the Fundamental Research Funds for the Central Universities of China under grant no. 2024CDJXY-022.}}

\author{Yu Sang$^{1}$
\quad Hai-Nan Lin$^{2,3;1)}$\email{linhn@cqu.edu.cn (Corresponding author)}
}

\maketitle
\hspace{1cm}

\address{$^1$ Center for Gravitation and Cosmology, College of Physical Science and Technology, Yangzhou University, Yangzhou 225009, China\\
$^2$ Department of Physics, Chongqing University, Chongqing 401331, China\\
$^3$ Chongqing Key Laboratory for Strongly Coupled Physics, Chongqing University, Chongqing 401331, China}

\begin{abstract}
The time series of energy and waiting time of magnetar bursts carry important information about the source activity. In this paper, we investigate the memory and dynamical stability of magnetar bursts from four soft gamma repeater (SGR) sources: SGR 1806$-$20, SGR 1900+14, SGR J1935+2154  and SGR J1550$-$5418. Based on the rescaled range analysis, we quantify the memory in magnetar bursts for the first time and find that there exists long-term memory in the time series of both waiting time and energy. We investigate the dynamical stability in the context of randomness and chaos. For all the four SGR samples, we find that the waiting time is not completely random, but the energy of two SGRs is consistent with a total random organization. Furthermore, both waiting time and energy exhibits weak chaos. We also find no significant difference between SGRs and repeating fast radio bursts (FRBs) in the randomness-chaos phase space. The statistical similarity between SGRs and repeating FRBs hints that there may be potential physical connection between these two phenomena.
\end{abstract}



\footnotetext[0]{\hspace*{-3mm}\raisebox{0.3ex}{$\scriptstyle\copyright$}2019
Chinese Physical Society and the Institute of High Energy Physics
of the Chinese Academy of Sciences and the Institute
of Modern Physics of the Chinese Academy of Sciences and IOP Publishing Ltd}%

\section{Introduction}\label{sec:intro}
Magnetars are neutron stars with extremely strong magnetic fields exceeding $\sim 10^{13} $ G \cite{Duncan1992,Mereghetti2008,Kaspi:2017fwg}. They have long rotational periods that typically last for several seconds and gradually spin down due to electromagnetic radiation. Magnetars are observationally recognized as soft gamma repeaters (SGRs), which persistently emit hard X-rays and soft gamma-rays \cite{Kouveliotou1998,Kouveliotou1999,Thompson2002}. The magnetar bursts usually have duration of  $\sim 0.1-1 ~\textrm{s}$ and peak luminosity in the range of $\sim 10^{39}-10^{41} ~\textrm{erg s}^{-1}$. Although it is widely accepted that the bursts are powered by the strong magnetic fields, the triggering mechanism remains unclear. Many theoretical models have been proposed to explain the triggering mechanism of SGRs, such as the crustquakes of neutron star \cite{Thompson:1995gw} and the magnetic reconnection \cite{Lyutikov:2003cz}.

There are already several studies on the statistical properties of SGRs, particularly focusing on the distributions of energy and waiting time \cite{Cheng1996,Gogus1999,Gogus2000,Chang:2017bnb,Cheng:2019ykn,Wei:2021kdw,Sang:2021cjq}. The cumulative energy distribution of 111 bursts from SGR 1806$-$20 was found to be well described by a power-law function with an index $\gamma = 1.66$, which is very close to the index $\gamma \approx 1.6$ of the earthquake Gutenberg–Richter power law \cite{Cheng1996}. 
Chang et al. \cite{Chang:2017bnb} investigated the cumulative distributions of SGR J1550$-$5418 and found that the distributions of fluence, peak flux and duration are well fitted by a bent power law, while the distribution of waiting time follows a simple power law. 
In addition to the power-law distribution of magnetars bursts, the fluctuations of bursts show scale-invariant properties \cite{Chang:2017bnb,Wei:2021kdw,Sang:2021cjq,Xiao:2024thj}. The probability density functions of fluctuations are well described by the Tsallis $q$-Gaussian function for fluence, peak flux, and duration of 384 bursts from SGR J1550$-$5418, and the $q$ values are consistent across different scale intervals, which indicates that the Tsallis $q$-Gaussian distribution of the fluctuations is scale-invariant \cite{Chang:2017bnb}.

The statistical similarity between SGRs and fast radio bursts (FRBs) has also been discussed \cite{Wei:2021kdw,Sang:2021cjq}, which is observationally motivated by the association between the Galactic FRB 200428 \cite{CHIMEFRB:2020abu,Bochenek:2020zxn} and the hard X-ray burst from SGR 1935+2154 \cite{Mereghetti:2020unm,Insight-HXMTTeam:2020dmu,Ridnaia:2020gcv,Tavani:2020adq,Younes:2020tac}. The repeating FRBs show properties similar to those of SGRs, including a power-law distribution for both energy and waiting time  \cite{Wang:2016lhy,Wang:2017agh,Wang:2019sio,Lin:2019ldn,Wang:2022gmu,Sang:2023zho}, as well as a scale-invariant Tsallis $q$-Gaussian distribution of their fluctuations \cite{Lin:2019ldn,Sang:2024swg,Wei:2021kdw,Wang:2022gmu,Gao:2024ekm}. The power-law distribution of energy and the scale-invariant fluctuations are predicted by the self-organized criticality (SOC) systems \cite{Bak:1987xua,aschwanden2011self}, which provides a potential explanation for the burst properties observed in both SGRs and repeating FRBs.

The statistical properties of the time series of repeating FRBs and SGRs have also been investigated recently \cite{Zhang:2023fmn,Yamasaki:2023fud,Sang:2024swg}. Particularly, Zhang et al. \cite{Zhang:2023fmn} employed the Pincus index and Lyapunov exponent to quantify the randomness and chaos of repeating FRB 20121102A and FRB 20190520B, and compared them with other natural phenomena such as pulsars, earthquakes, solar flares and Brownian motion. The repeating FRBs were found to exhibit high randomness and low chaos, mimicking the behavior of Brownian motion in the randomness-chaos phase space. The same dynamical stability analysis was also applied to magnetar bursts from SGR J1550-5418 and SGR J1935+2154. It was found that these magnetar bursts have a distinct separation from FRBs in time domain, but has no significant difference compared to FRBs in energy domain \cite{Yamasaki:2023fud}. Another interesting phenomenon identified through the analysis of time series is the long-term memory, which has been observed in repeating FRBs \cite{Wang:2023sjs,Wang:2023wcb,Sang:2024swg}, but has not yet been studied in magnetar bursts.

In this paper, we investigate the statistical properties of magnetar burst using four active SGRs, which consists of approximately 2000 bursts in total. We particularly focus on the long-term memory and dynamical stability of SGRs. The structure of this paper is arranged as follows: The dataset used in this paper is presented in Section {\ref{sec:data}}. The long-term memory is studied in Section {\ref{sec:memory}}. The randomness and chaos are analyzed in Section {\ref{sec:randomness}}. Finally we give the discussion and conclusions in Section {\ref{sec:conclusion}}.

\section{Data Samples}\label{sec:data}

\begin{figure*}[htbp]
    \centering
	\includegraphics[width=0.45\textwidth]{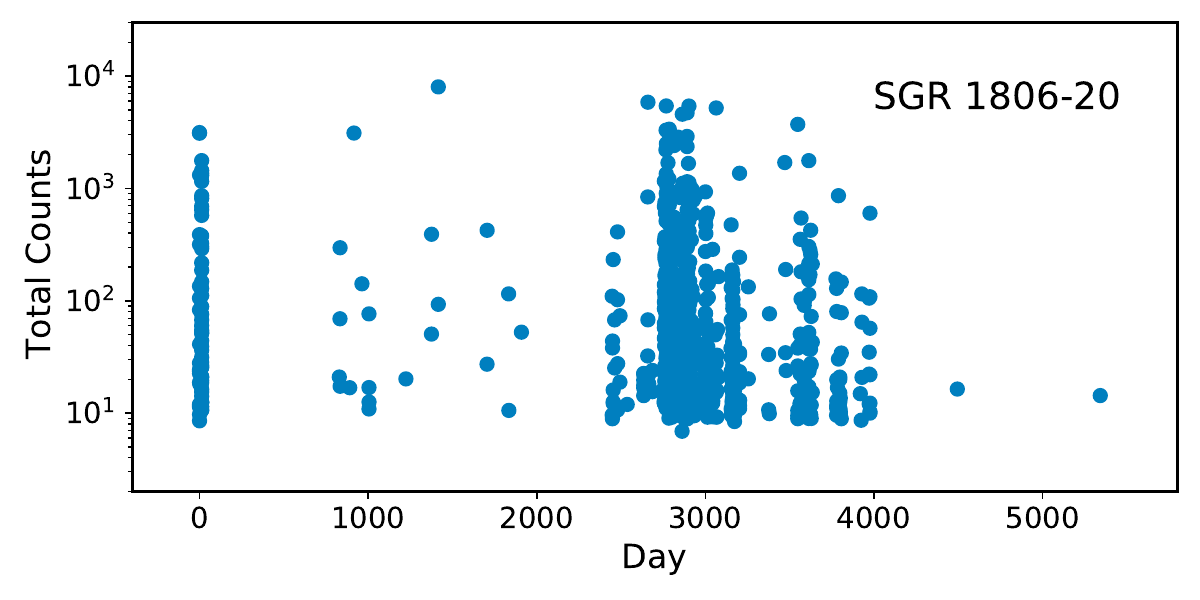}
    \includegraphics[width=0.45\textwidth]{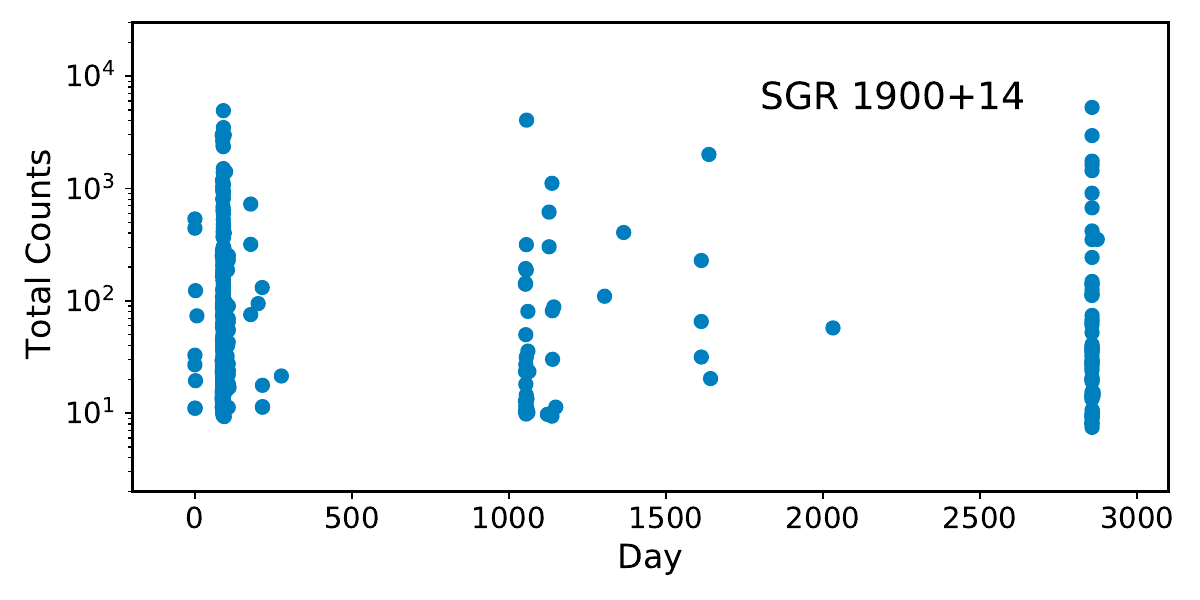}
    \includegraphics[width=0.45\textwidth]{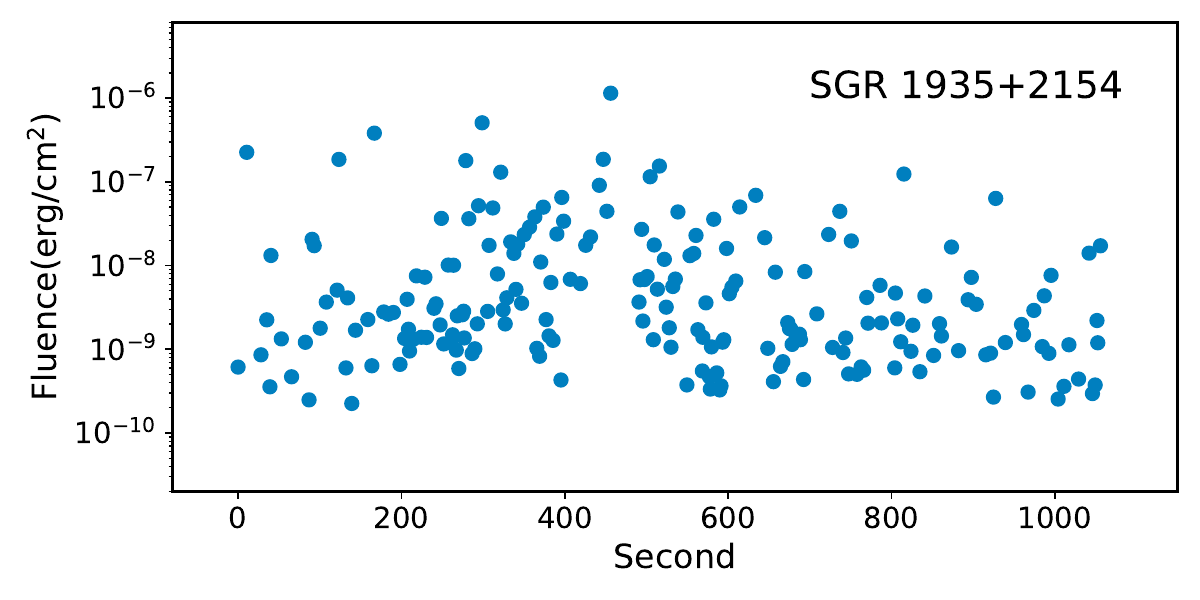}
    \includegraphics[width=0.45\textwidth]{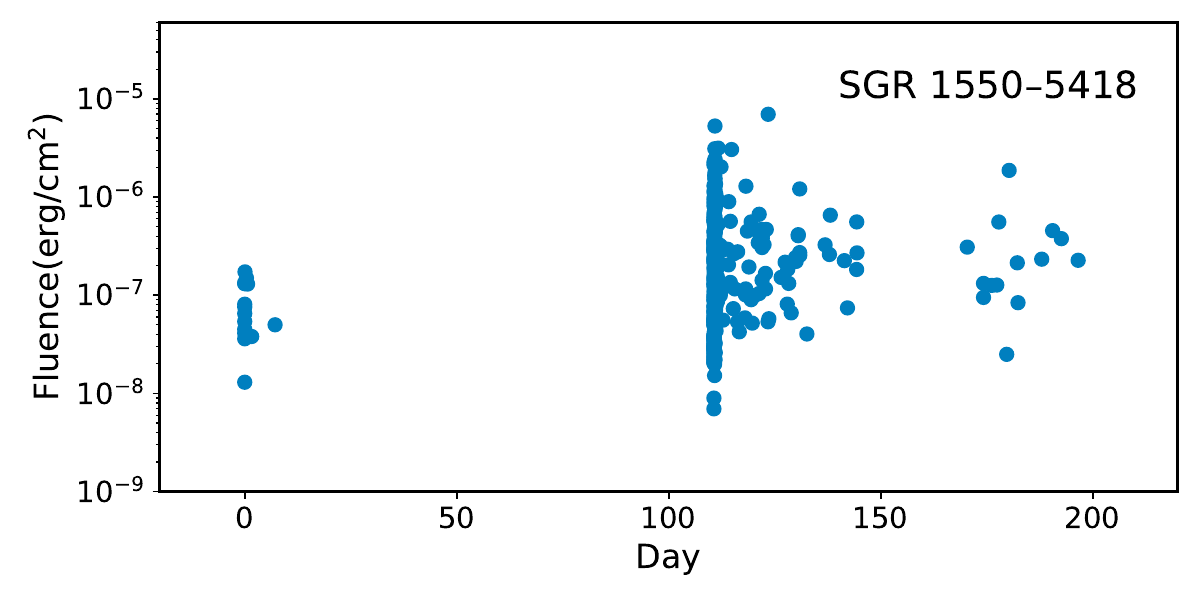}
    \caption{Photon counts or fluence vs. the arrival time for all the four SGR samples. The arrival time of the first burst is set to zero.}
    \label{fig:energytime}
\end{figure*}

In the analysis, we use the burst data from four SGR sources: SGR 1806$-$20, SGR 1900+14, SGR J1935+2154, and SGR J1550$-$5418. The first sample consists of 924 bursts from SGR 1806$-$20 detected by the Rossi X-ray Timing Explorer (RXTE) between 1996 and 2011. The second sample consists of 432 bursts from SGR 1900+14 detected by RXTE between 1998 and 2006. Both samples are available at the online database\footnote{http://magnetars.sabanciuniv.edu/} constructed by the high-energy astrophysics group at Sabancı University. The burst time and total count of each bursts can be found in the database. The third sample is 217 bursts from SGR J1935+2154 observed by the NICER telescope during the 1120s burst storm period on 2020 April 28 \cite{Younes:2020hie}. The database provides the burst start time, duration and flux of the bursts. We can also calculate the fluence by the product of duration and the time-averaged flux. The fourth sample is 384 bursts from SGR J1550-5418 observed in the three active episodes in 2008-2009 by the Gamma-ray Burst Monitor onboard the Fermi Gamma-ray Space Telescope (\textit{Fermi}/GBM) \cite{Collazzi:2015kea}. The $T_{\rm 90}$ start time and fluence are given in the database. All the four samples are shortly summarized in Table \ref{tab:sample}. The photon counts or fluence as a function of arrival time are depicted in Figure \ref{fig:energytime}.

\begin{table}[htbp]
\centering
\caption{The SGR samples used in our analysis.}
\label{tab:sample}
\begin{tabular}{lllll}
\hline\hline 
SGRs & $N$ & detector & Date & Reference\\
\hline
SGR 1806$-$20 & 924 & RXTE & 1996-2011 & \href{http://magnetars.sabanciuniv.edu/}{Online Database}\\
SGR 1900$+$14 & 432 & RXTE & 1998-2006 & \href{http://magnetars.sabanciuniv.edu/}{Online Database}\\
SGR J1935$+$2154  & 217 & NICER & 2020 & \cite{Younes:2020hie}\\
SGR J1550$-$5418 & 384 & GBM & 2008-2009 & \cite{Collazzi:2015kea} \\
\hline
\end{tabular}
\end{table}

In this paper, we investigate the long-term memory and dynamical stability of magnetar bursts through analyzing the time sequence of waiting time and energy. The waiting time is defined as the time difference between two successive bursts, $\Delta T = T_{i+1} - T_i$. We use the burst time of SGR 1806$-$20 and SGR 1900+14, the burst start time of SGR J1935+2154, and the $T_{\rm 90}$ start time of SGR J1550$-$5418 as the arrival time of the $i$-th burst $T_i$. Note that each sample consists of different observing sessions. The duration of an observing session is expected to be no more than a half of the orbital periods of satellites. The orbital periods of satellites around the earth are 93, 93 and 95 minutes for RXTE, NICER and Fermi satellites, respectively. Therefore, we use a uniform criterion to discard the waiting time larger than 1 hour to avoid the long observing gaps. The burst energy is not a directly observable quantity. For a specific SGR source, the burst energy is proportional to the fluence or the photon counts. Rescaling the burst energy by a universal constant does not affect the results we discussed bellow. Hence we directly use the fluence or the photon counts as the representation of burst energy. Specifically, we use the time sequence of total count in SGR 1806$-$20 and SGR 1900+14 samples, and the time sequence of fluence (flux multiplied by duration) in SGR J1935+2154 and SGR J1550$-$5418 samples as the energy sequence in the following analysis.

\section{Long-term Memory}\label{sec:memory}

The long-term memory observed in the time series data is a fascinating phenomenon which has already been identified in several natural events, such as earthquakes \cite{barani2018long}, solar flares \cite{Aschwanden_2021} and repeating FRBs \cite{Wang:2023sjs,Wang:2023wcb,Sang:2024swg}. In this section, we present the measurement of long-term memory in magnetar bursts for the first time. We employ the rescaled range analysis (R/S analysis) to calculate the Hurst exponent $H$ \cite{hurst1956problem,hurst1957suggested}, which is a quantification of the long-term memory in time series data. 
If the rescaled range analysis yields \( H > 0.5 \), it indicates the presence of positive long-range correlations in the time series data. Conversely, if \( H < 0.5 \), negative long-range correlations are presented. If \( H = 0.5 \), it suggests an absence of long-range correlations within the data.

Here we give a concise introduction of the rescaled range analysis method \cite{Mandelbrot1969Robustness,Weron_2002,MERAZ2022126631}. 
A time series of length $N$ is divided into $l$ non-overlapping subseries, each of length $n$.
For each subseries $X_m$, where $m=1,2,...,l$, the rescaled range analysis follows the steps bellow: (a) find the mean value $E_m$ and the standard deviation $S_m$ for the subseries $X_{m}$; (b) normalize the data in subseries $X_{m}$ by subtracting the mean value to obtain a mean-adjusted time series, $Y_{i,m}=X_{i,m}-E_m$ for $i=1,...,n$; (c) construct a cumulative deviation time series given by $Z_{i,m}=\sum_{j=1}^i Y_{j,m}$ for $i=1,...,n$; (d) determine the series range $R_m = \max \{Z_{1,m},...,Z_{n,m}\} - \min \{Z_{1,m},...,Z_{n,m}\}$; and (e) rescale the range using the standard deviation, $R_m/S_m$. Finally we calculate the mean value of the rescaled range for all subseries of length $n$, 
\begin{equation}
(R/S)_n  = \frac{1}{l}\sum_{m=1}^l R_m/S_m.
\end{equation}
By varying the length $n$ of subseries, we construct a series of the rescaled range, which is asymptotically following the relation
\begin{equation}
(R/S)_n = Cn^H,
\end{equation}
where $H$ is the Hurst exponent, which can be derived through linear regression method in the logarithmic scale,
\begin{equation}
\ln (R/S)_n = \ln C + H \ln n.
\end{equation}

\begin{figure*}[htbp]
    \centering
	\includegraphics[width=0.45\textwidth]{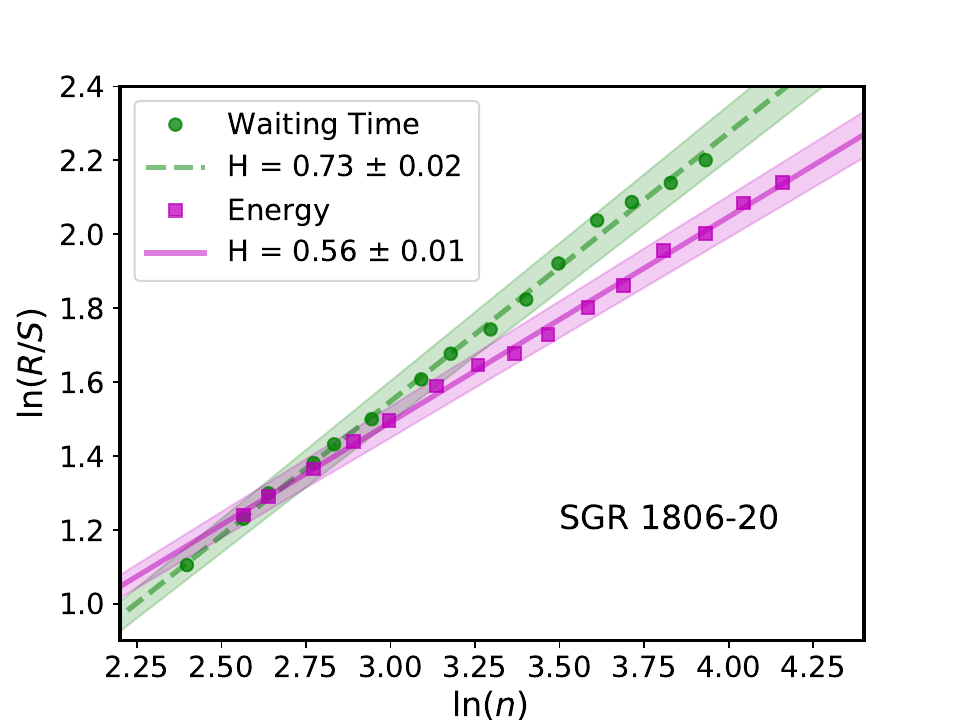}
    \includegraphics[width=0.45\textwidth]{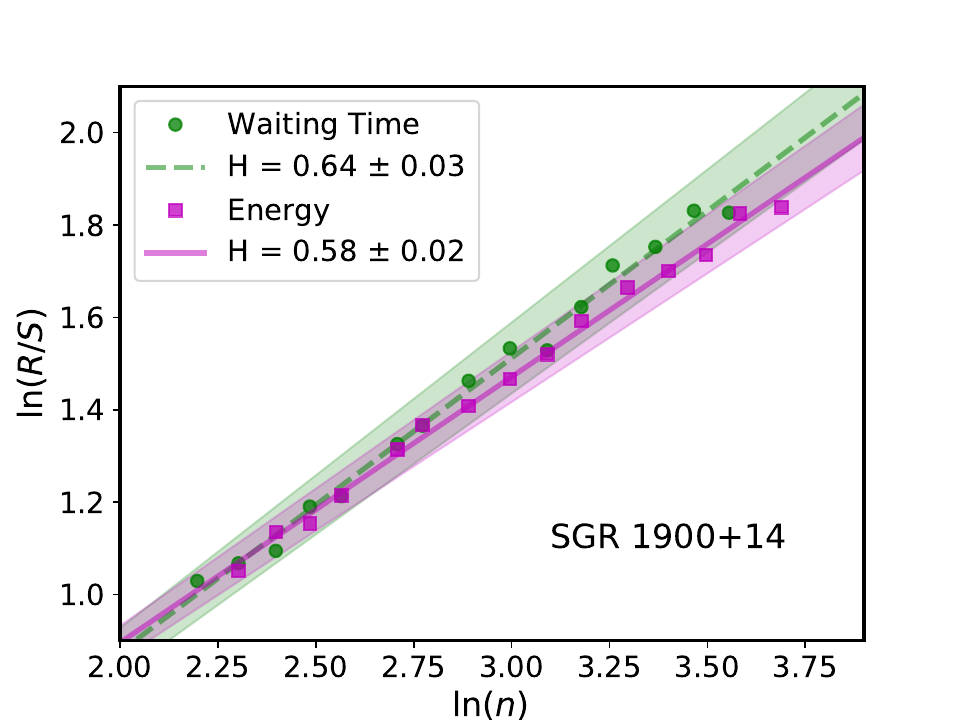}
    \includegraphics[width=0.45\textwidth]{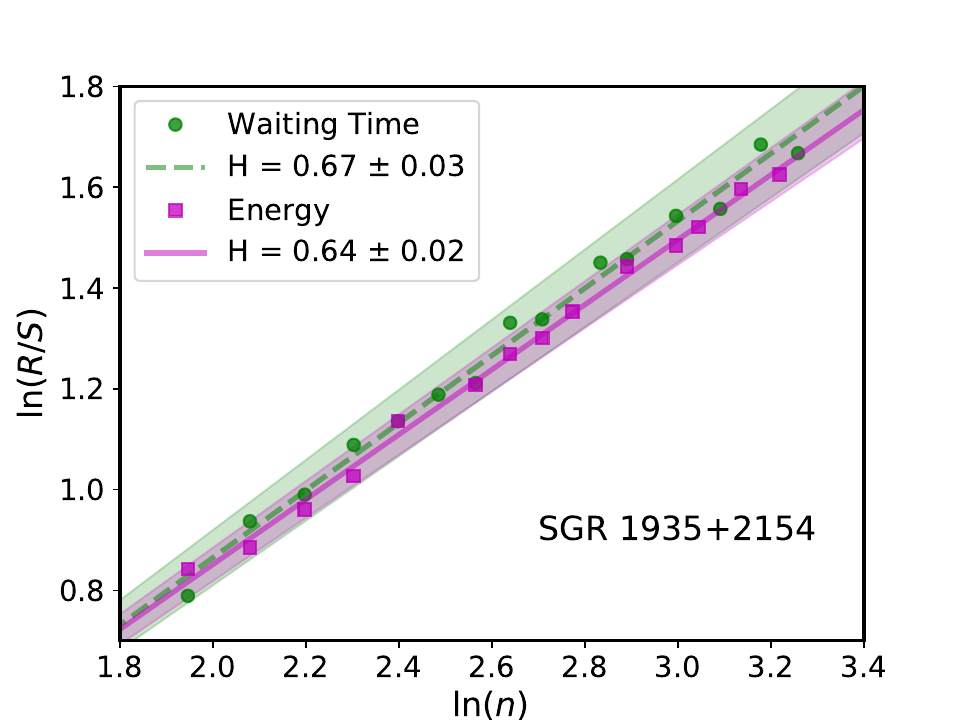}
    \includegraphics[width=0.45\textwidth]{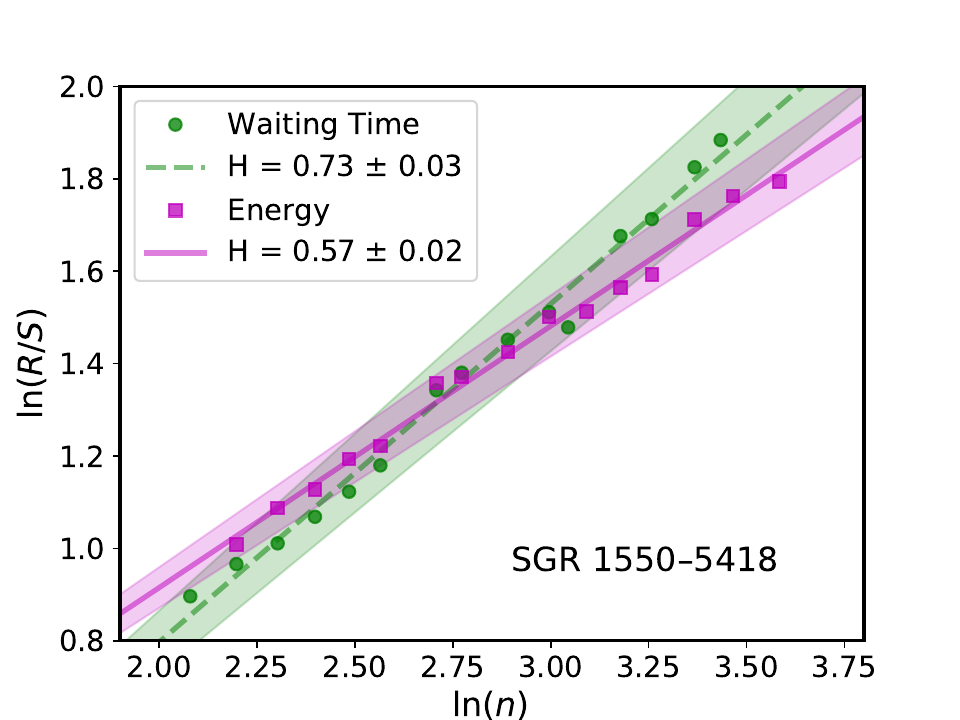}
    \caption{The rescaled range series of waiting time and energy for SGRs. The dashed lines and solid lines are the linear regression curves of waiting time and energy, respectively. The shaded regions represent the $1\sigma$ uncertainty of the linear regression.
    }
    \label{fig:hurst}
\end{figure*}

In this paper, we employ the R/S analysis on the time series of waiting time and energy of magnetar bursts. The public package \textsf{NOLDS} \cite{scholzel_2020_3814723} is used in the analysis. Figure \ref{fig:hurst} illustrates the rescaled range series as a function of $n$ for four SGR samples, with waiting time represented in green dots and energy in magenta squares. The points correspond to the rescaled range series derived from the sample data and the straight lines represent the results of the linear regression. As we expect, the rescaled range series is a simple power-law function of $n$. The data points fit well with a straight line in the log-log plot, where the slope of the line gives the Hurst exponent $H$. The best-fitting lines obtained using the usual least-$\chi^2$ method are shown in Figure \ref{fig:hurst}, where the shaded regions represent the $1\sigma$ uncertainty. The Hurst exponents $H$ for waiting time and energy of the four SGR samples are summarized in Table \ref{tab:hurst}. We also calculate the intrinsic scatter $\sigma_\mathrm{int}$, which is defined by the square root of the reduced chi-square. For the samples of SGR 1806$-$20, SGR 1900+14, SGR J1935+2154 and SGR J1550$-$5418, the Hurst exponents are $H=0.73 \pm 0.02$, $0.64 \pm 0.03$, $0.67 \pm 0.03$ and $0.73 \pm 0.03$ for waiting time, and $H=0.56 \pm 0.01$, $0.58 \pm 0.02$, $0.64 \pm 0.02$ and $0.57 \pm 0.02$ for energy, respectively. The Hurst exponents are larger than $0.5$ at $\gtrsim 3\sigma$ confidence level for all the samples. Therefore, we conclude that long-term memory exists in the time series of both waiting time and energy for the magnetar bursts.

\begin{table}[htbp]
\centering
\caption{The Hurst exponent $H$ of waiting time and energy for SGRs. The uncertainty is the intrinsic scatter of the linear regression.}
\label{tab:hurst}
\begin{tabular}{lll} 
\hline\hline 
 & Waiting time  & Energy \\
\hline
SGR 1806$-$20  & 0.73 $\pm$ 0.02 & 0.56 $\pm$ 0.01 \\
SGR 1900+14 & 0.64 $\pm$ 0.03 & 0.58 $\pm$ 0.02 \\
SGR J1935+2154 & 0.67 $\pm$ 0.03 & 0.64 $\pm$ 0.02 \\
SGR J1550$-$5418 & 0.73 $\pm$ 0.03 & 0.57 $\pm$ 0.02 \\
\hline
\end{tabular}
\end{table}

\section{Randomness and Chaos}\label{sec:randomness}

In this section, we explore the dynamical stability of magnetar bursts in the context of randomness and chaos. The randomness and chaos of a time series are quantified by the Pincus index (PI) \cite{Pincus1991App} and the largest Lyapunov exponent (LLE) \cite{WOLF1985285}, respectively. We compute the PI and LLE values for the waiting time and energy in four SGRs samples and illustrate the results in randomness-chaos phase space.

The PI value is a quantification of the randomness in a dynamical system, and its calculation is based on the concept of approximate entropy. The approximate entropy of a time series ${u(i)}$ with length $N$ is defined by \cite{Pincus1991App,e21060541}
\begin{align}
   \nonumber
    {\rm ApEn}(m,r,N) \simeq & -\frac{1}{N-m} \\ 
   &\times \sum_{i=1}^{N-m}\log \frac{\sum_{j=1}^{N-m}\theta({\rm dist}[x_{m+1}(j), x_{m+1}(i)]-r)}{\sum_{j=1}^{N-m} \theta({\rm dist}[x_{m}(j), x_{m}(i)]-r)}. 
\end{align}
Here $x_{m}(i)=[u(i), ..., u(i + m - 1)]$ and $x_{m}(j)=[u(j), ..., u(j+m-1)]$ are the subseries of ${u(i)}$, and the embedding dimension $m$ is the length of the subseries. ${\rm dist}[x,y]$ is the Chebyshev distance between $x$ and $y$. $\theta(x)$ is the step function, i.e. $\theta=1$ for $x\geq 0$, and $\theta=0$ for $x<0$. We use the public code \textsf{EntropyHub} \cite{EntropyHub} to compute ApEn in this paper. We follow the traditional convention to take the embedding dimension $m=2$ and vary the distance threshold $r$ in the range of [0.01,0.09] multiplied by the standard deviation of the data series. The maximum approximate entropy (MAE) \cite{delgado2019quantifying,e21060541} is defined as the maximum value of ApEn across varying the value of $r$,
\begin{equation}
    {\rm MAE} = \max_r  [{\rm ApEn}(m,r,N) ].
\end{equation}
Through bootstrap sampling we can compute PI for a series, which is defined as the ratio of MAE of the original series to that of the randomly shuffled series,
\begin{equation}
    {\rm PI} = \frac{{\rm MAE}_\textrm{original}}{{\rm MAE}_\textrm{shuffled}}.
\end{equation}

The PI quantify the randomness of a time series, with a value of zero indicating a completely ordered system and a value of unity representing complete randomness. In this paper, we perform random shuffling on the original series 1000 times, so obtain 1000 values of ${\rm MAE_{shuffled}}$. The bootstrap sampling also gives the uncertainty of PI, which is propagated from the uncertainty of ${\rm MAE_{shuffled}}$,
\begin{equation}
\frac{\sigma_{\rm PI}}{\rm PI} = \frac{\sigma_{{\rm MAE}_\textrm{shuffled}}}{{\rm MAE}_\textrm{shuffled}},
\end{equation}
where $\sigma_{{\rm MAE}_\textrm{shuffled}}$ is defined as the standard deviation of the ${\rm MAE}$ values of the randomly shuffled series.

\begin{figure*}[htbp]
    \centering
	\includegraphics[width=0.45\textwidth]{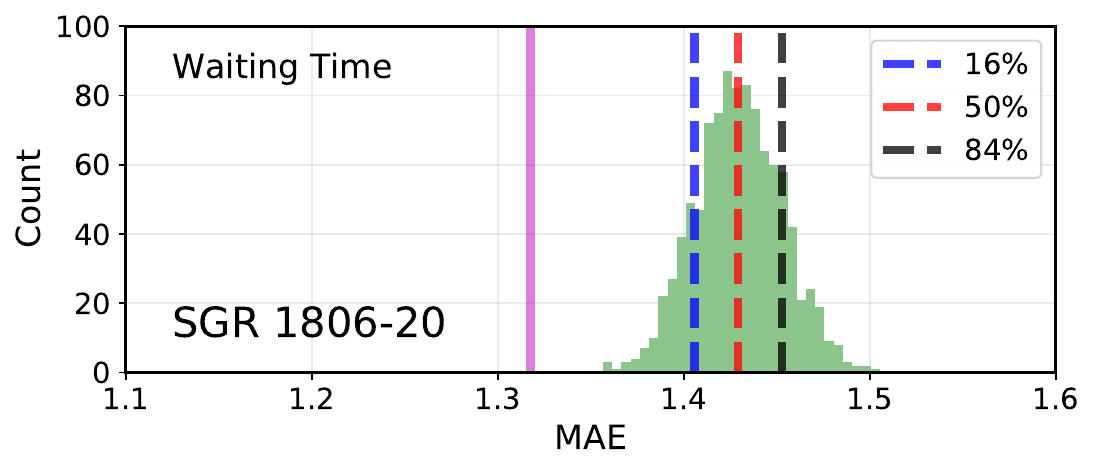}
    \includegraphics[width=0.45\textwidth]{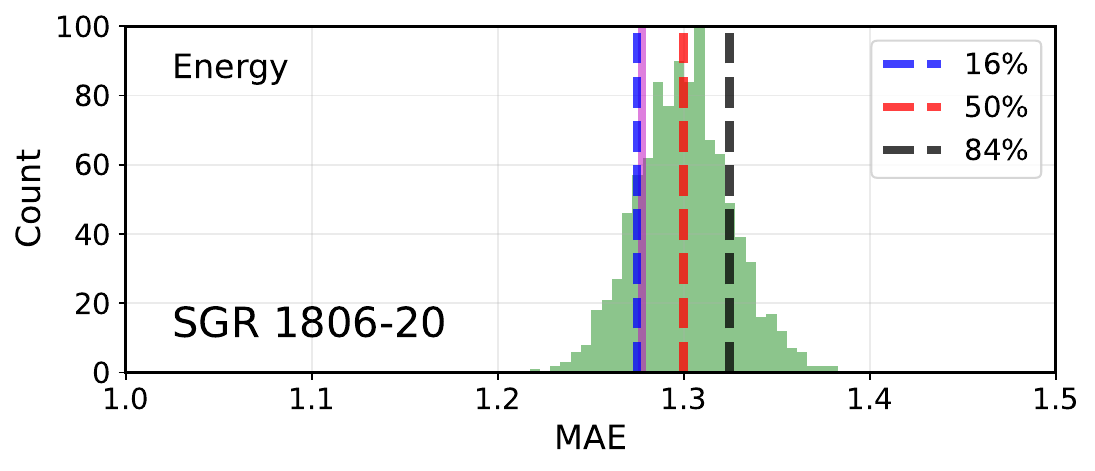}
    \includegraphics[width=0.45\textwidth]{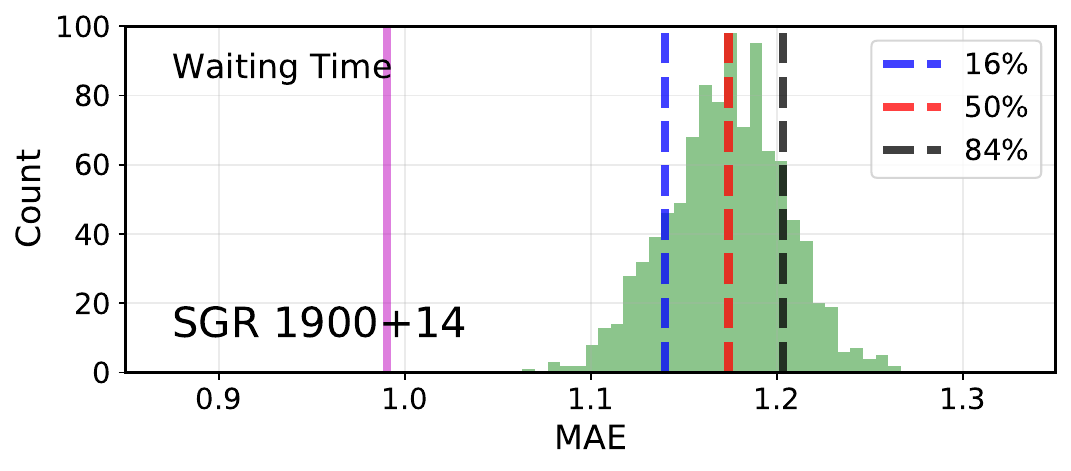}
    \includegraphics[width=0.45\textwidth]{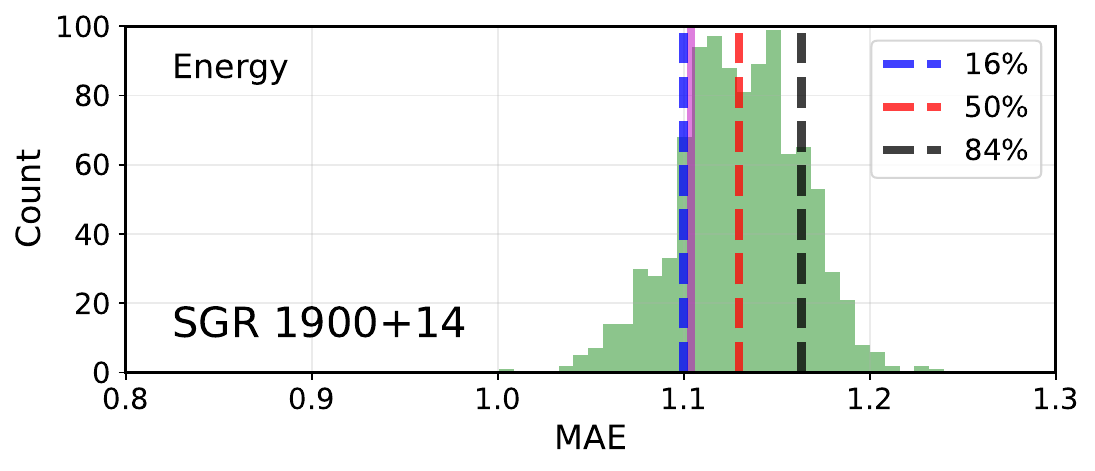}
    \includegraphics[width=0.45\textwidth]{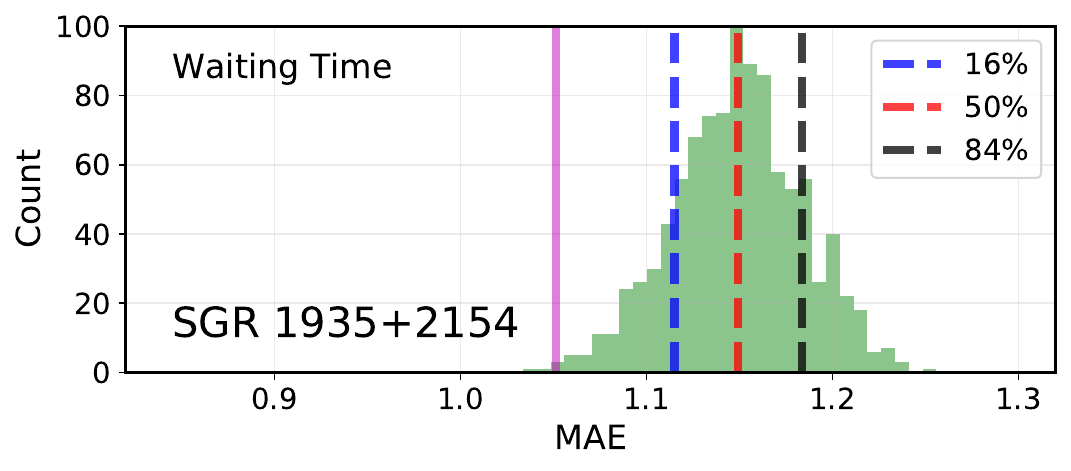}
    \includegraphics[width=0.45\textwidth]{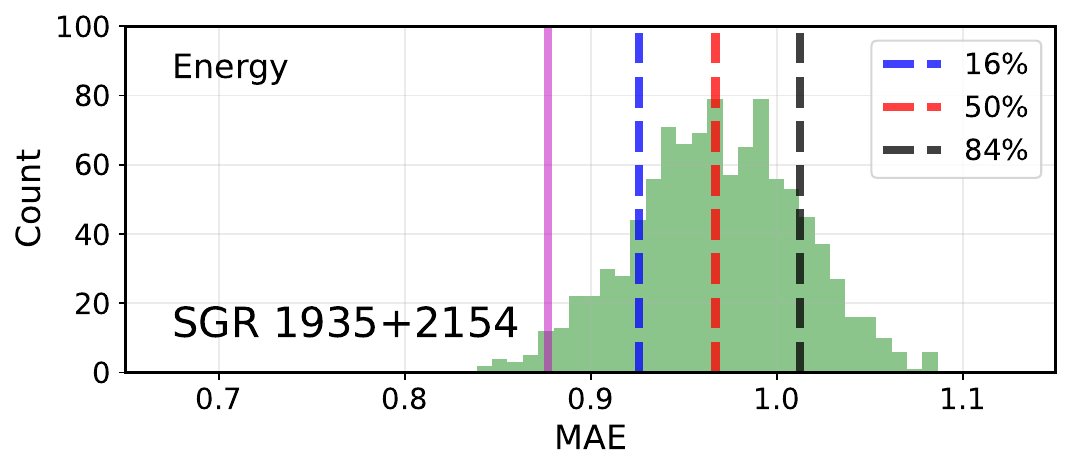}
    \includegraphics[width=0.45\textwidth]{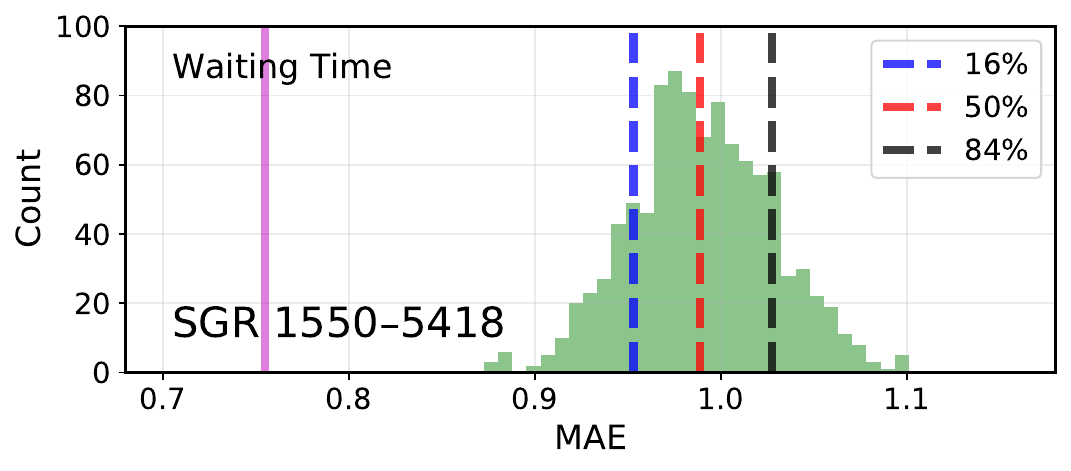}
    \includegraphics[width=0.45\textwidth]{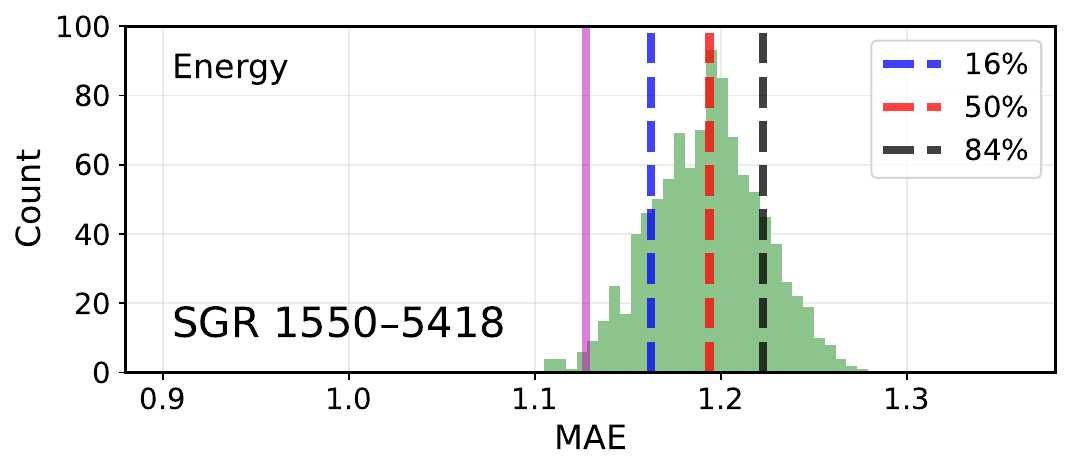}
    \caption{ The MAE distribution for SGRs. The magenta solid line is MAE of the original series. The green histogram is the distribution of MAE of the 1000 shuffled series. The blue, red and black dashed lines are the 16th, 50th, and 84th percentile of the 1000 simulations, respectively.}
    \label{fig:pi}
\end{figure*}

We calculate the MAE and PI values for the time series of waiting time and energy for the four SGR samples.
In Figure \ref{fig:pi}, we show the MAE value for the original series in magenta solid line, and the distribution of the MAE values for the 1000 shuffled series in green histogram. The 16th, 50th and 84th percentiles of the distribution are plotted in blue, red and black dashed lines, respectively. In Table \ref{tab:pi}, we list the PI values and their uncertainties. We can quantify the randomness of the samples through the distance between ${\rm MAE}_\textrm{original}$ and ${\rm MAE}_\textrm{shuffled}$, or directly through the values of PI. As can be seen, the waiting time series of SGR 1900+14 and SGR J1550$-$5418 significantly deviate from a random organization, with PI values of $0.84 \pm 0.02$ and $0.76 \pm 0.03$, respectively. For SGR 1806$-$20 and J1935+2154, the waiting time series are also slightly different from a totally random organization. Hence, we conclude that the waiting time series are not totally random for all the four SGRs samples. As for the energy series, the PI value of a totally random organization (${\rm PI}=1$) is inside the $1 \sigma$ confidence interval of the PI values of SGR 1806$-$20 and SGR 1900+14. In contrast, the PI values of SGR J1935+2154 and SGR J1550$-$5418 deviates from complete randomness (${\rm PI}=1$) at more than $2 \sigma$ confidence level. Therefore, the energy series is totally random for SGR 1806$-$20 and SGR 1900+14 samples, but it is less random for SGR J1935+2154 and SGR J1550$-$5418 samples.

\begin{table}[htbp]
\centering
\caption{The PI values of waiting time and energy for SGRs. The uncertainty is the standard deviation of the randomly shuffled series.}
\label{tab:pi}
\begin{tabular}{lcccc} 
\hline\hline 
 & Waiting time  & Energy \\
\hline
SGR 1806$-$20 & 0.92 $\pm$ 0.02 & 0.98 $\pm$ 0.02 \\
SGR 1900+14 & 0.84 $\pm$ 0.02 & 0.98 $\pm$ 0.03 \\
SGR J1935+2154 & 0.92 $\pm$ 0.03 & 0.91 $\pm$ 0.04 \\
SGR J1550$-$5418 & 0.76 $\pm$ 0.03 & 0.95 $\pm$ 0.02 \\
\hline
\end{tabular}
\end{table}

The LLE is a quantification of chaos in a non-linear dynamical system, and it quantifies the local stability features of attractors and other invariant sets in phase space. In the $m-$dimensional phase space, an initially infinitesimal $m$-sphere of radius $r_0$ will deform into a $m$-ellipsoid due to the locally deforming nature of the phase flow. The length of the $i$-th ellipsoidal principal axis evolves exponentially with time $r_i=r_0\exp(\lambda_it)$. The Lyapunov exponents is then defined by \cite{WOLF1985285}
\begin{equation}
 \lambda_i=\lim_{t\rightarrow\infty}\frac{1}{t}\ln\frac{r_i}{r_0},~~~~i=1,2,\cdots,m.
\end{equation}
LLE is the maximum value of Lyapunov exponents $\lambda_i$, which quantifies the rate of separation of two adjacent trajectories in phase space. A positive LLE value means that two infinitely close trajectories in phase space will diverge exponentially as time elapses, thus implies the existence of chaos. On the contrary, a negative value of LLE suggests a stable system. We use the algorithm of Rosenstein et al. \cite{ROSENSTEIN1993117} implemented in the public package \textsf{NOLDS} \cite{scholzel_2020_3814723} to calculate LLE, where the embedding dimension is chosen to be the default value $m=10$. Since the LLE is the maximum value of the whole spectrum of Lyapunov exponents, it is hard to define the uncertainty. One possible way is to calculate the LLE values by varying the embedding dimension $m$, and use the standard deviation of LLE values to represent the uncertainty. However, we find that the LLE is almost independent of the embedding dimension (see bellow). So we ignore the uncertainty of LLE here. Table \ref{tab:lyap} summarizes the LLE values of waiting time and energy series in the four SGRs. Although all the LLE values are positive, they are very close to zero. Therefore, we conclude that there is no strong evidence for the existence of chaos for all the four SGR samples.

\begin{table}[htbp]
\centering
\caption{The LLE values of waiting time and energy for SGRs.}
\label{tab:lyap}
\begin{tabular}{lllllll} 
\hline\hline 
 & Waiting time  & Energy \\
\hline
SGR 1806$-$20 & 0.051 & 0.074 \\
SGR 1900+14 & 0.056 & 0.079 \\
SGR J1935+2154 & 0.034 & 0.061 \\
SGR J1550$-$5418 & 0.137 &  0.046 \\
\hline
\end{tabular}
\end{table}

\begin{figure*}[htbp]
    \centering
	\includegraphics[width=0.45\textwidth]{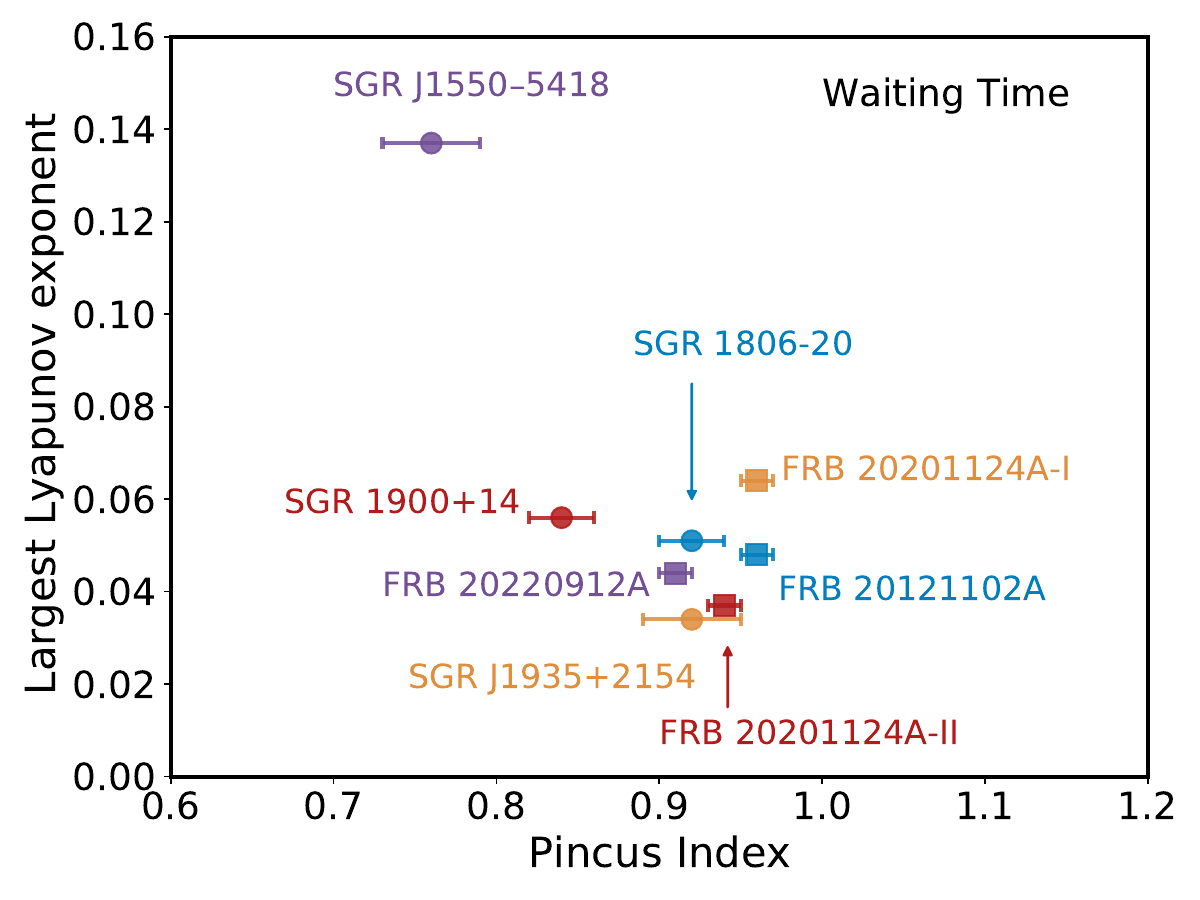}
    \includegraphics[width=0.45\textwidth]{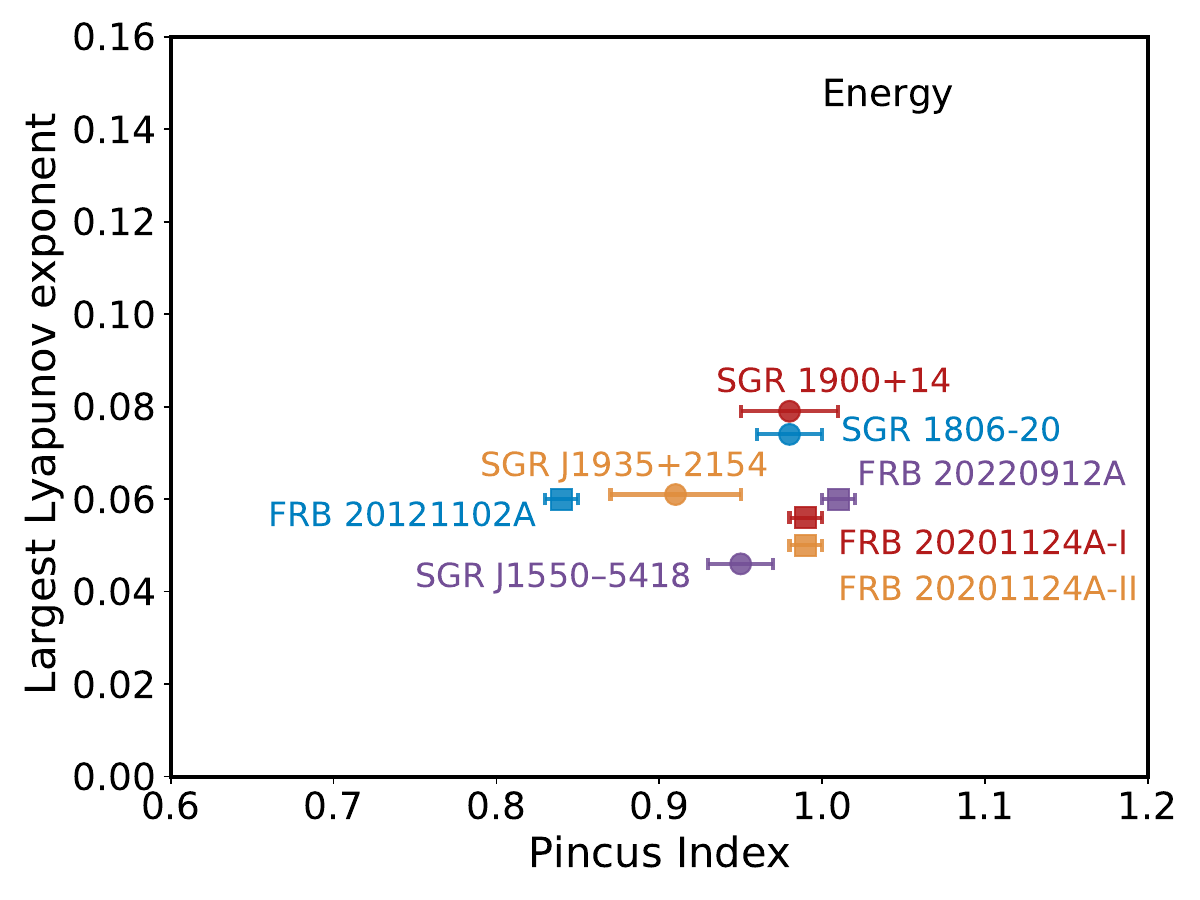}
    \caption{The randomness-chaos plane of waiting time (left-panel) and energy (right panel) for SGRs and FRBs. The dots and squares represent for SGRs and FRBs, respectively.}
    \label{fig:lle_pi}
\end{figure*}

In Figure \ref{fig:lle_pi}, we plot the LLE and PI values for each SGR in the randomness-chaos phase plane. As a comparison, we also illustrate the LLE and PI values of four samples from three extremely active repeating FRBs observed by the FAST telescope, i.e. FRB 20121102A, FRB 20201124A and FRB 20220912A. The bursts from FRB 20201124A are divided into two samples, which are observed in two separate periods with about three months observational gap. The PI values of the four FRB samples were calculated in our previous work \cite{Sang:2024swg}, and the LLE values are calculated by reanalyzing the same data samples using the method as applied in this paper. In the time domain (the left panel of Figure \ref{fig:lle_pi}), except for SGR J1550$-$5418, the rest three SGR samples and all the four FRB samples concentrate on a small region (with the average values ${\rm PI}\sim 0.9$ and ${\rm LLE}\sim 0.05$). SGR J1550$-$5418 seems to be an outlier in the randomness-chaos phase plane, it has a smaller PI ($\sim 0.76$) and a larger LLE ($\sim 0.14$) than the other SGRs and FRBs. Hence the waiting time of SGR J1550$-$5418 are less random but more chaotic than other samples. In the energy domain (the right panel of Figure \ref{fig:lle_pi}), all the data samples seem to concentrate on a small region in the randomness-chaos phase plane (with the average values ${\rm PI}\sim 0.9$ and ${\rm LLE}\sim 0.06$). Especially, the two samples from FRB 20201124A do not significantly diverge in the randomness-chaos plane, implying that there is no strong temporal variability in the burst activity. This result is consistent with our previous conclusion \cite{Sang:2024swg}. In summary, we conclude that there is no significant difference between SGRs and FRBs in the randomness-chaos phase plane, for both waiting time and energy.

\begin{figure*}[htbp]
    \centering
	\includegraphics[width=0.45\textwidth]{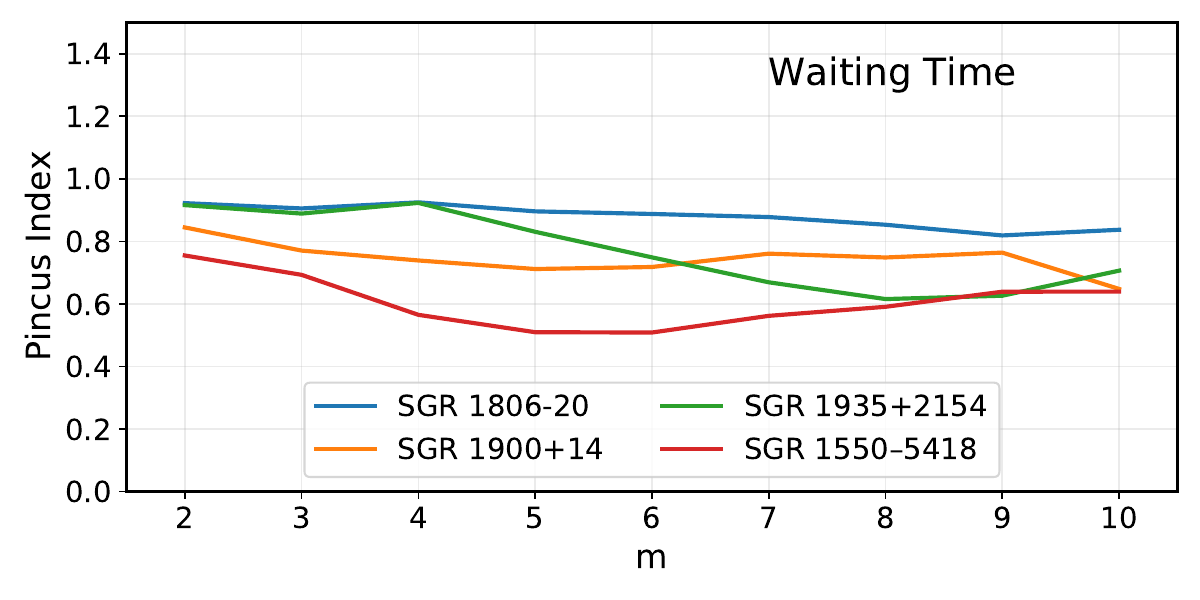}
    \includegraphics[width=0.45\textwidth]{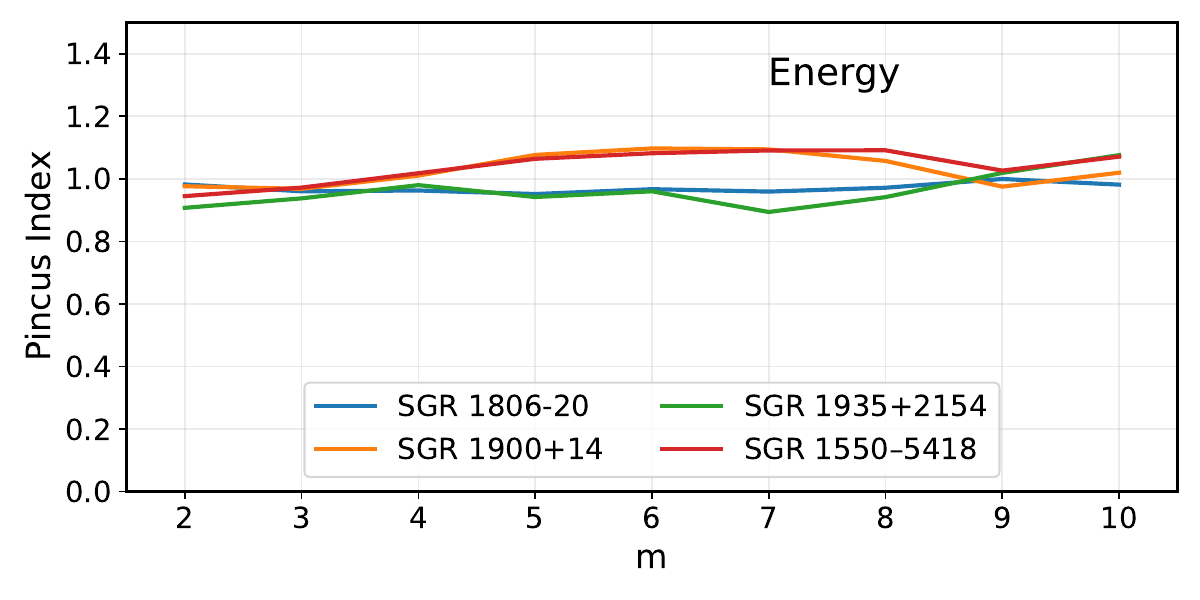}
    \includegraphics[width=0.45\textwidth]{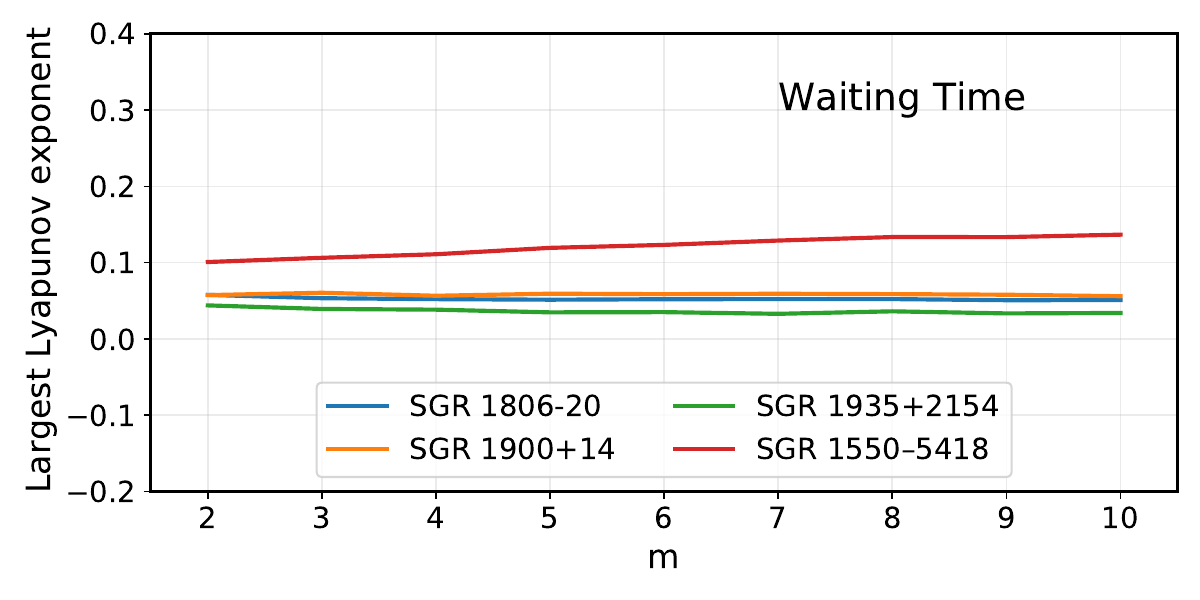}
    \includegraphics[width=0.45\textwidth]{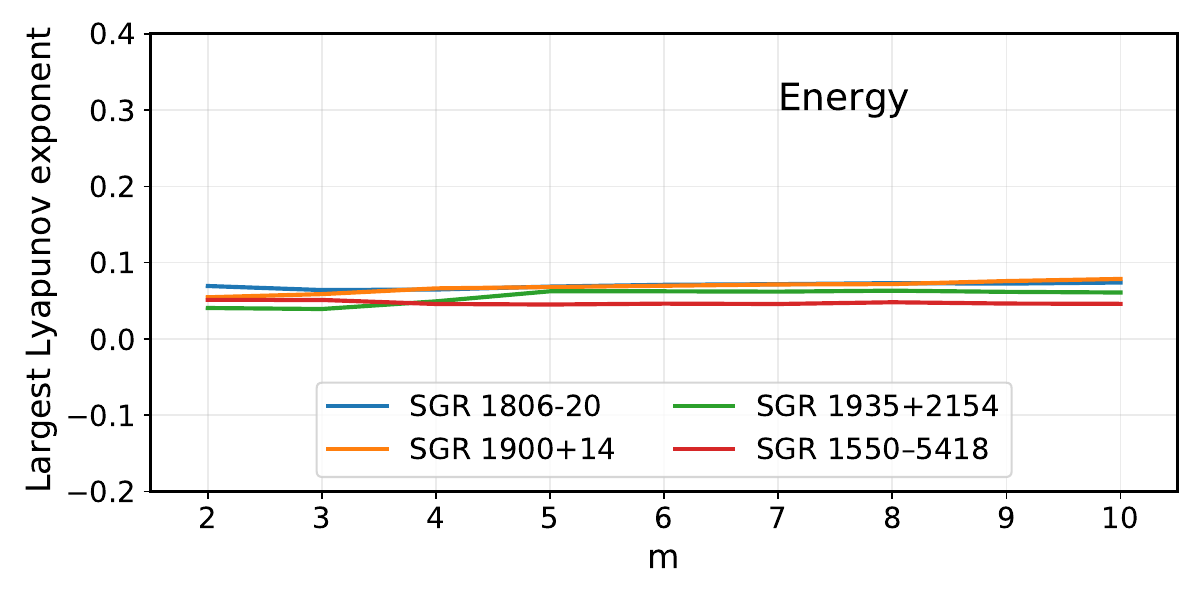}
    \caption{The PI and LLE values vs. the embedding dimension $m$.}
    \label{fig:PI_m}
\end{figure*}

Both the PI and LLE values depend on the embedding dimension $m$. We use different embedding dimension $m$ in the calculations of PI and LLE to assess the robustness of our results. The PI and LLE values as a function of $m$ are depicted in Figure \ref{fig:PI_m} for waiting time and energy of the four SGR samples. As can be seen, both PI and LLE values do not significantly affected by the choice of embedding dimension, implying the robustness of our results.

\clearpage
\section{Discussions and Conclusions}\label{sec:conclusion}

In this paper, we investigated the statistical properties of waiting time and energy of magnetar bursts using four samples. Using the rescaled range analysis to calculate the Hurst exponent, we measure the long-term memory in magnetar bursts for the first time, and found that long-term memory exists in the time series of both waiting time and energy of magnetar bursts. We investigated the dynamical stability of magnetar bursts through the measurement of randomness and chaos, which are quantified by the Pincus index and the largest Lyapunov exponent, respectively. In the time domain, all the four SGR samples show evidence for the deviation from random organization. In the energy domain, however, the result is somewhat subtle: SGR 1806$-$20 and SGR 1900+14 is totally consistent with random organization, but SGR J1935+2154 and SGR J1550$-$5418 seem to be less random. We also found that both waiting time and energy exhibit weak chaos in all the four SGR samples. Finally, we compared SGRs with repeating FRBs and found that there is no significant difference between these two astronomical phenomena in the randomness-chaos phase space, for both waiting time and energy.

Previous discussions on the statistical similarity between SGRs and repeating FRBs focus on the power-law distribution of the observed quantities such as energy, waiting time and burst duration, or the scale-invariant Tsallis $q$-Gaussian distribution of their fluctuations. Here we extend the scope of the investigation to the long-term memory and randomness-chaos characteristics, and we also found that SGRs share similar properties with repeating FRBs. The Hurst exponent $H$ for repeating FRBs are found to be around 0.6 \cite{Sang:2024swg}, consistent with the $H$ values for SGRs obtained in this paper. Therefore the long-range correlations in SGRs and FRBs are positive but not strong. The PI values for FRBs and SGRs are also very similar, ${\rm PI}\sim 0.8-1.0$. These statistical similarities, along with the observation of the FRB-associated magnetar SGR 1935+2154 \cite{CHIMEFRB:2020abu,Bochenek:2020zxn}, imply that there may be common emission mechanism between SGRs and FRBs.

The repeating FRBs were also compared with other physical phenomena in the randomness-chaos phase space. For instance, Zhang et al. \cite{Zhang:2023fmn} found FRBs deviate significantly from pulsars, earthquakes, and solar flares. We found that, for both waiting time and energy, SGRs do not deviate significantly from FRBs within the randomness-chaos phase space. The randomness and chaos have also been investigated for two SGRs by Yamasaki et al. \cite{Yamasaki:2023fud}. In their work, the authors compared SGRs and FRBs in the randomness-chaos plane, and found that SGRs exhibit significantly lower randomness and a slightly higher degree of chaos compared to FRBs in time domain, but they exhibit a broad consistency in the energy domain. Our results, however, show that SGRs and FRBs are consistent with each other in both time domain and energy domain, which conflict with the results of Yamasaki et al. \cite{Yamasaki:2023fud}. One reason may be the different data samples we used. Another reason may be that Yamasaki et al. \cite{Yamasaki:2023fud} used the energy fluctuation in the analysis, while we used the energy itself. This issue needs to be further investigated with a larger data sample in the future.

\vspace{5mm}
\centerline{\rule{80mm}{0.5pt}}
\vspace{2mm}

\bibliographystyle{cpc}
\bibliography{reference}

\begin{thebibliography}{10}
\providecommand{\url}[1]{\texttt{#1}}
\providecommand{\urlprefix}{URL }
\providecommand{\eprint}[2][]{\url{#2}}

\bibitem{Duncan1992}
R.~C. Duncan and C.~Thompson.
\newblock The Astrophysical Journal, \textbf{392}: L9--L13 (1992).

\bibitem{Mereghetti2008}
S.~Mereghetti.
\newblock The Astronomy and Astrophysics Review, \textbf{15}~(4): 225--287
  (2008).

\bibitem{Kaspi:2017fwg}
V.~M. Kaspi and A.~Beloborodov.
\newblock Ann. Rev. Astron. Astrophys., \textbf{55}: 261--301 (2017).

\bibitem{Kouveliotou1998}
C.~Kouveliotou, S.~Dieters, T.~Strohmayer \emph{et~al.}
\newblock Nature, \textbf{393}~(6682): 235--237 (1998).

\bibitem{Kouveliotou1999}
C.~Kouveliotou, T.~Strohmayer, K.~Hurley \emph{et~al.}
\newblock The Astrophysical Journal, \textbf{510}~(2): L115 (1999).

\bibitem{Thompson2002}
C.~Thompson, M.~Lyutikov, and S.~Kulkarni.
\newblock The Astrophysical Journal, \textbf{574}: 332--355 (2002).

\bibitem{Thompson:1995gw}
C.~Thompson and R.~C. Duncan.
\newblock Mon. Not. Roy. Astron. Soc., \textbf{275}: 255--300 (1995).

\bibitem{Lyutikov:2003cz}
M.~Lyutikov.
\newblock Mon. Not. Roy. Astron. Soc., \textbf{346}: 540 (2003).

\bibitem{Cheng1996}
B.~Cheng, R.~I. Epstein, R.~A. Guyer \emph{et~al.}
\newblock Nature, \textbf{382}~(6591): 518--520 (1996).

\bibitem{Gogus1999}
E.~G\"{o}\v{g}\"{u}\c{s}, P.~M. Woods, C.~Kouveliotou \emph{et~al.}
\newblock The Astrophysical Journal Letters, \textbf{526}~(2): L93 (1999).

\bibitem{Gogus2000}
E.~G\"{o}\v{g}\"{u}\c{s}, P.~M. Woods, C.~Kouveliotou \emph{et~al.}
\newblock The Astrophysical Journal Letters, \textbf{532}~(2): L121 (2000).

\bibitem{Chang:2017bnb}
Z.~Chang, H.-N. Lin, Y.~Sang \emph{et~al.}
\newblock Chin. Phys. C, \textbf{41}~(6): 065104 (2017).

\bibitem{Cheng:2019ykn}
Y.~Cheng, G.~Q. Zhang, and F.~Y. Wang.
\newblock Mon. Not. Roy. Astron. Soc., \textbf{491}~(1): 1498--1505 (2020).

\bibitem{Wei:2021kdw}
J.-J. Wei, X.-F. Wu, Z.-G. Dai \emph{et~al.}
\newblock Astrophys. J., \textbf{920}~(2): 153 (2021).

\bibitem{Sang:2021cjq}
Y.~Sang and H.-N. Lin.
\newblock Mon. Not. Roy. Astron. Soc., \textbf{510}~(2): 1801--1808 (2022).

\bibitem{Xiao:2024thj}
S.~Xiao, S.-N. Zhang, S.-L. Xiong \emph{et~al.}
\newblock Mon. Not. Roy. Astron. Soc., \textbf{528}~(2): 1388--1392 (2024).

\bibitem{CHIMEFRB:2020abu}
B.~C. Andersen \emph{et~al.} (CHIME/FRB).
\newblock Nature, \textbf{587}~(7832): 54--58 (2020).

\bibitem{Bochenek:2020zxn}
C.~D. Bochenek, V.~Ravi, K.~V. Belov \emph{et~al.}
\newblock Nature, \textbf{587}~(7832): 59--62 (2020).

\bibitem{Mereghetti:2020unm}
S.~Mereghetti \emph{et~al.}
\newblock Astrophys. J. Lett., \textbf{898}~(2): L29 (2020).

\bibitem{Insight-HXMTTeam:2020dmu}
C.~K. Li \emph{et~al.} (Insight-HXMT Team).
\newblock Nature Astronomy, \textbf{5}: 378--384 (2021).

\bibitem{Ridnaia:2020gcv}
A.~Ridnaia \emph{et~al.}
\newblock Nature Astron., \textbf{5}~(4): 372--377 (2021).

\bibitem{Tavani:2020adq}
M.~Tavani \emph{et~al.}
\newblock Nature Astron., \textbf{5}~(4): 401--407 (2021).

\bibitem{Younes:2020tac}
G.~Younes \emph{et~al.}
\newblock Nature Astron., \textbf{5}~(4): 408--413 (2021).

\bibitem{Wang:2016lhy}
F.~Y. Wang and H.~Yu.
\newblock JCAP, \textbf{03}: 023 (2017).

\bibitem{Wang:2017agh}
W.~Wang, R.~Luo, H.~Yue \emph{et~al.}
\newblock Astrophys. J., \textbf{852}~(2): 140 (2018).

\bibitem{Wang:2019sio}
F.~Y. Wang and G.~Q. Zhang.
\newblock The Astrophysical Journal, \textbf{882}~(2): 108 (2019).

\bibitem{Lin:2019ldn}
H.-N. Lin and Y.~Sang.
\newblock Mon. Not. Roy. Astron. Soc., \textbf{491}~(2): 2156--2161 (2020).

\bibitem{Wang:2022gmu}
Z.-H. Wang, Y.~Sang, and X.~Zhang.
\newblock Res. Astron. Astrophys., \textbf{23}~(2): 025002 (2023).

\bibitem{Sang:2023zho}
Y.~Sang and H.-N. Lin.
\newblock Mon. Not. Roy. Astron. Soc., \textbf{523}~(4): 5430--5441 (2023).

\bibitem{Sang:2024swg}
Y.~Sang and H.-N. Lin.
\newblock Mon. Not. Roy. Astron. Soc., \textbf{533}~(1): 872--879 (2024).

\bibitem{Gao:2024ekm}
C.-Y. Gao and J.-J. Wei.
\newblock Astrophys. J., \textbf{968}~(1): 40 (2024).

\bibitem{Bak:1987xua}
P.~Bak, C.~Tang, and K.~Wiesenfeld.
\newblock Phys. Rev. Lett., \textbf{59}: 381--384 (1987).

\bibitem{aschwanden2011self}
M.~Aschwanden.
\newblock \emph{Self-Organized Criticality in Astrophysics: The Statistics of
  Nonlinear Processes in the Universe}.
\newblock Springer Praxis Books,  (Springer Berlin Heidelberg2011).

\bibitem{Zhang:2023fmn}
Y.-K. Zhang, D.~Li, Y.~Feng \emph{et~al.}
\newblock Sci. Bull., \textbf{69}: 1020--1026 (2024).

\bibitem{Yamasaki:2023fud}
S.~Yamasaki, E.~Gogus, and T.~Hashimoto.
\newblock Mon. Not. Roy. Astron. Soc., \textbf{528}~(1): L133--L138 (2023).

\bibitem{Wang:2023sjs}
P.~Wang \emph{et~al.}
\newblock Astrophys. J., \textbf{975}~(2): 188 (2024).

\bibitem{Wang:2023wcb}
F.~Y. Wang, Q.~Wu, and Z.~G. Dai.
\newblock Astrophys. J. Lett., \textbf{949}~(2): L33 (2023).

\bibitem{Younes:2020hie}
G.~Younes \emph{et~al.}
\newblock Astrophys. J. Lett., \textbf{904}~(2): L21 (2020).
\newblock [Erratum: Astrophys.J.Lett. 913, L17 (2021)].

\bibitem{Collazzi:2015kea}
A.~C. Collazzi \emph{et~al.}
\newblock Astrophys. J. Suppl., \textbf{218}~(1): 11 (2015).

\bibitem{barani2018long}
S.~Barani, C.~Mascandola, E.~Riccomagno \emph{et~al.}
\newblock Scientific reports, \textbf{8}~(1): 5326 (2018).

\bibitem{Aschwanden_2021}
M.~J. Aschwanden and J.~R. Johnson.
\newblock The Astrophysical Journal, \textbf{921}~(1): 82 (2021).

\bibitem{hurst1956problem}
H.~E. Hurst.
\newblock Hydrological Sciences Journal, \textbf{1}~(3): 13--27 (1956).

\bibitem{hurst1957suggested}
H.~E. Hurst.
\newblock Nature, \textbf{180}~(4584): 494--494 (1957).

\bibitem{Mandelbrot1969Robustness}
B.~B. Mandelbrot and J.~R. Wallis.
\newblock Water Resources Research, \textbf{5}~(5): 967--988 (1969).

\bibitem{Weron_2002}
R.~Weron.
\newblock Physica A: Statistical Mechanics and its Applications,
  \textbf{312}~(1–2): 285–299 (2002).

\bibitem{MERAZ2022126631}
M.~Meraz, J.~Alvarez-Ramirez, and E.~Rodriguez.
\newblock Physica A: Statistical Mechanics and its Applications, \textbf{589}:
  126631 (2022).

\bibitem{scholzel_2020_3814723}
C.~Schölzel.
\newblock Nonlinear measures for dynamical systems (2020).

\bibitem{Pincus1991App}
S.~M. Pincus.
\newblock Proceedings of the National Academy of Sciences, \textbf{88}~(6):
  2297--2301 (1991).

\bibitem{WOLF1985285}
A.~Wolf, J.~B. Swift, H.~L. Swinney \emph{et~al.}
\newblock Physica D: Nonlinear Phenomena, \textbf{16}~(3): 285--317 (1985).

\bibitem{e21060541}
A.~Delgado-Bonal and A.~Marshak.
\newblock Entropy, \textbf{21}~(6) (2019).

\bibitem{EntropyHub}
M.~W. Flood and B.~Grimm.
\newblock PLOS ONE, \textbf{16}~(11): 1--20 (2021).

\bibitem{delgado2019quantifying}
A.~Delgado-Bonal.
\newblock Scientific reports, \textbf{9}~(1): 12761 (2019).

\bibitem{ROSENSTEIN1993117}
M.~T. Rosenstein, J.~J. Collins, and C.~J. {De Luca}.
\newblock Physica D: Nonlinear Phenomena, \textbf{65}~(1): 117--134 (1993).

\end{thebibliography}

\end{document}